\documentclass[sigconf,screen]{acmart}

\usepackage{enumitem}
\usepackage{amsmath,amsfonts}
\usepackage{algorithmic}
\usepackage{array}
\usepackage[caption=false,font=normalsize,labelfont=sf,textfont=sf]{subfig}
\usepackage{textcomp}
\usepackage{stfloats}
\usepackage{url}
\usepackage{verbatim}
\usepackage{graphicx}
\usepackage{balance}
\usepackage{adjustbox}
\usepackage{booktabs}
\usepackage{multirow}
\usepackage{threeparttable}
\usepackage{arydshln}
\usepackage{tcolorbox}
\usepackage{soul}
\usepackage{fontawesome}
\usepackage{makecell}
\usepackage{natbib}
\usepackage{siunitx}
\usepackage{microtype}
\usepackage{colortbl}
\definecolor{mygray}{gray}{.9}
\usepackage{tikz}
\usepackage[T1]{fontenc}
\usepackage{aecompl}
\usepackage{xparse}

\AtBeginDocument{
  \providecommand\BibTeX{{
    \normalfont B\kern-0.5em{\scshape i\kern-0.25em b}\kern-0.8em\TeX}}}

\newcommand{\tech}{\textit{Specine}}
\newcommand{\techWoPTC}{\textit{Specine$_{woPTC}$}}
\newcommand{\techWoT}{\textit{Specine$_{woT}$}}
\newcommand{\techWoA}{\textit{Specine$_{woA}$}}
\newcommand{\techWTF}{\textit{Specine$_{wTF}$}}
\newcommand{\techWoAR}{\textit{Specine$_{woAR}$}}

\newcommand{\revision}[1]{{\color{black}{#1}}}
\newcommand{\Comment}[1]{}

\NewDocumentCommand{\framecolorbox}{oommm}
 {% #1 = width (optional)
  \IfValueTF{#1}
   {\IfValueTF{#2}
    {\fcolorbox{#3}{#4}{\makebox[#1][#2]{#5}}}
    {\fcolorbox{#3}{#4}{\makebox[#1]{#5}}}%
   }
   {\fcolorbox{#3}{#4}{#5}}%
 }

\copyrightyear{2026}
\acmYear{2026}
\setcopyright{rightsretained}
\acmConference[ICSE '26]{2026 IEEE/ACM 48th International Conference on Software Engineering}{April 12--18, 2026}{Rio de Janeiro, Brazil}
\acmBooktitle{2026 IEEE/ACM 48th International Conference on Software Engineering (ICSE '26), April 12--18, 2026, Rio de Janeiro, Brazil}
\acmPrice{}
\acmDOI{10.1145/3744916.3764572}
\acmISBN{979-8-4007-2025-3/26/04}
\begin{document}

\title{Aligning Requirement for Large Language Model's Code Generation}

\author{Zhao Tian}
\orcid{0000-0002-9316-7250}
\affiliation{
  \institution{College of Intelligence and \\
  Computing, Tianjin University}
  \city{Tianjin}
  \country{China}
}
\email{tianzhao@tju.edu.cn}

\author{Junjie Chen}
\authornote{Junjie Chen is the corresponding author.}
\orcid{0000-0003-3056-9962}
\affiliation{
  \institution{College of Intelligence and \\
  Computing, Tianjin University}
  \city{Tianjin}
  \country{China}
}
\email{junjiechen@tju.edu.cn}

\begin{abstract}
Code generation refers to the automatic generation of source code based on a given programming specification, which has garnered significant attention particularly with the advancement of large language models (LLMs). 
However, due to the inherent complexity of real-world problems, the LLM-generated code often fails to fully align with the provided specification. 
While state-of-the-art agent-based techniques have been proposed to enhance LLM code generation, they overlook the critical issue of specification perception, resulting in persistent misalignment issues.
Given that accurate perception of programming specifications serves as the foundation of the LLM-based code generation paradigm, ensuring specification alignment is particularly crucial.
In this work, we draw on software requirements engineering to propose \tech{}, a novel specification alignment technique for LLM code generation.
Its key idea is to identify misaligned input specifications, lift LLM-perceived specifications, and align them to enhance the code generation performance of LLMs.
Our comprehensive experiments on four state-of-the-art LLMs across five challenging competitive benchmarks by comparing with ten state-of-the-art baselines, demonstrate the effectiveness of \tech{}. 
For example, \tech{} outperforms the most effective baseline, achieving an average improvement of 29.60\% across all subjects in terms of Pass@1.
\end{abstract}




\maketitle

\section{Introduction}
\label{sec:introduction}
Code generation refers to the automatic generation of source code from a given programming specification. 
In recent years, it has attracted significant attention from both academia and industry due to its potential to reduce repetitive programming tasks and improve software development productivity~\cite{peng2023impact,kazemitabaar2023studying,tian2022learning,gao2025trae}. 
Advancements in LLMs, such as DeepSeek-Coder~\cite{guo2025deepseek} and Gemini~\cite{team2024gemini}, have led to substantial improvements in code generation~\cite{yang2024exploring}. 
However, LLMs still face significant challenges, particularly when handling complex requirements~\cite{zhang2024paircoder,mu2024clarifygpt,tian2025fixing}. 
For example, even advanced GPT-4~\cite{achiam2023gpt} generates code passing all test cases for only 12.10\% of real-world competitive programming problems\cite{islam2024mapcoder}. 
These performance issues impede the practical application of LLMs, potentially compromising software quality~\cite{tian2023code,tian2023fly}. 
Therefore, enhancing the code generation performance of LLMs is critical.

Recently, various prompt-based~\cite{li2025structured,olausson2024selfrepair,tian2025fixing} and agent-based~\cite{dong2024selfcollaboration,hong2023metagpt,zhang2024paircoder} techniques have been proposed to enhance the code generation performance of LLMs. 
In general, these techniques share four main steps:
(1) Perception, where the LLM is provided with a requirement specification and perceives its intent;
(2) Planning, where the LLM employs diverse planing strategies (e.g., task decomposition or algorithm selection) based on its specification perception;
(3) Implementation, where the LLM generates code based on the devised plan;
(4) Repair, where the LLM leverages feedback (e.g., test execution messages) to fix incorrect code.
Among them, agent-based code generation has emerged as the state-of-the-art~\cite{huang2023agentcoder,lin2025flowgen}, which utilizes specialized LLM agents to simulate the human software development workflow.
For example, Self-collaboration~\cite{dong2024selfcollaboration} enables LLMs to take on different roles (e.g., analyst, coder, and tester), each responsible for a specific sub-task in the development process. 
PairCoder~\cite{zhang2024paircoder}, the state-of-the-art agent-based technique, employs a navigator agent to explore diverse solution plans and a driver agent to repair implementations using execution feedback.

However, existing LLM-based code generation techniques~\cite{li2025structured,olausson2024selfrepair,dong2024selfcollaboration,hong2023metagpt} mainly focus on the latter three stages (i.e., planning, implementation, and repair), while overlooking a fundamental issue in the first stage: the perception bias of LLMs towards the original input specification. 
Specifically, LLMs may implicitly ignore or misperceive key ingredients within the input specification, leading to generated code that deviates from the intended requirement. 
\textit{The difference between the original input specification and the actual LLM-perceived specification is referred to as specification misalignment.} 
Actually, inaccurate perception can limit the capability of optimizing the subsequent three stages in improving LLM-based code generation, since it may fundamentally produce inaccuracies in the three stages (such as inaccurate planning).
In addition, software requirements engineering research~\cite{pohl1996requirements,zave1997four,macaulay2012requirements,chen2025deep} emphasizes that effective software development must be built upon an accurate perception of fundamental specifications. 
Therefore, within the current LLM-driven automated code generation paradigm, ensuring specification alignment is particularly critical. 
That is, addressing the specification misalignment in LLM-based code generation is a key direction.

In this paper, we propose \textbf{\tech{}} (\textbf{Spec}ification Al\textbf{i}g\textbf{n}m\textbf{e}nt), a novel specification alignment technique to enhance the code generation performance of LLMs. 
Specifically, \tech{} automatically identifies the misaligned specification, lifts the LLM-perceived specification, and aligns it with the original input specification, thereby facilitating correct code generation.
However, designing an effective specification alignment technique presents the following key challenges:
\textit{(1) How to identify the misaligned ones from input specifications?}
Performing alignment on all cases can incur unnecessary overhead, even may misprocess the originally-correct cases.
\textit{(2) How to obtain the actual LLM-perceived specification for the misaligned one?}
Once a misaligned specification is identified, explicitly extracting the actual LLM-perceived specification facilitates the subsequent specification alignment. 
\textit{(3) How to align the LLM-perceived specification to generate correct code?} 
Designing comprehensive and effective alignment rules is essential for aligning the LLM-perceived specification with the original input specification, thereby enhancing code generation performance.

To address the first challenge, \tech{} implements a dual-agent component comprising a coder agent to generate the initial code and a tester agent to estimate its correctness.  
The correctness of the generated code serves as an indicator of the LLM's perception of the input specification~\cite{wei2024requirements,guo2025intention}.
If the specification is identified as potentially misaligned, \tech{} activates subsequent components for specification alignment.
To address the second challenge, \tech{} employs a novel specification lifting component that explicitly extracts the LLM-perceived specification for effective misalignment analysis.  
This LLM-perceived specification is lifted from the low-level generated code using a domain-specific language (DSL) of requirement specifications, providing a high-level standardized representation.
To address the third challenge, \tech{} applies ten pre-defined alignment rules (derived from software requirements engineering~\cite{greenspan1994formal,glinz2000problems,doe2011recommended}) to systematically align different ingredients of the misaligned specification.
In this way, \tech{} generates an aligned specification that is accurately perceived by the LLM, ultimately enhancing its code generation performance.

We conduct extensive experiments to evaluate \tech{} on four advanced LLMs (i.e., DeepSeek-Coder-1.5~\cite{guo2024deepseek}, Qwen2.5-Coder~\cite{hui2024qwen2}, GPT-4o-mini~\cite{openai2024introducing}, and Gemini-1.5-Flash~\cite{team2024gemini}) based on five challenging competitive program-solving benchmarks (i.e., APPS~\cite{hendrycks2021measuring}, APPS-Eval~\cite{dong2024codescore}, CodeContests-Raw~\cite{li2022competition}, CodeContests~\cite{li2022competition}, and xCodeEval~\cite{khan2023xcodeeval}).
Our results demonstrate that \tech{} significantly outperforms all 10 popular or state-of-the-art baselines across all 20 subjects (4 LLMs $\times$ 5 benchmarks), demonstrating our idea for enhancing LLM-based code generation by specification alignment.
For example, the average improvement of \tech{} over all 10 baselines is 29.60\%$\sim$93.55\% in terms of Pass@1 (measuring the ratio of programming problems for which the generated code passes all test cases) across all subjects.
Particularly, on the APPS dataset, the best performance in terms of Pass@1 achieved by all the four LLMs with the 10 baselines is 55.67\% (using the Gemini-1.5-Flash model with the TGen technique), but \tech{} achieves 65.33\% based on the same LLM.
Also, we investigate the influence of the number of iterations, a key hyper-parameter in the agent-based framework.
Our results show that as the number of iterations increases, \tech{} consistently outperforms all baselines.
Furthermore, we construct five variants of \tech{} for an ablation study, confirming the contribution of each main component in \tech{}.

The main contributions of this paper are summarized as follows:
\begin{itemize}[leftmargin=10pt]
    \item \textbf{Novel Perspective}: We propose a novel perspective for enhancing the code generation performance of LLMs through the sophisticated specification alignment.
    
    \item \textbf{Tool Implementation}: We implement \tech{} following the novel perspective, which consists of identifying misaligned input specifications, lifting LLM-perceived specifications, and aligning them to generate the correct code.

    \item \textbf{Performance Evaluation}: We conduct extensive experiments on four advanced LLMs across five challenging benchmarks by comparing with ten state-of-the-art baselines, demonstrating the effectiveness of \tech{} in improving LLMs' code generation. 
    
    \item \textbf{Data Availability}: We publicly release all accessible dataset and our source code at the project homepage~\cite{homepage2025} to facilitate the experiment replication, future research, and practical adoption.

\end{itemize}

\section{Motivating Example}
\label{sec:motivating_example}

\begin{figure}[t!]
    \centering
    \includegraphics[width=1.0\linewidth]{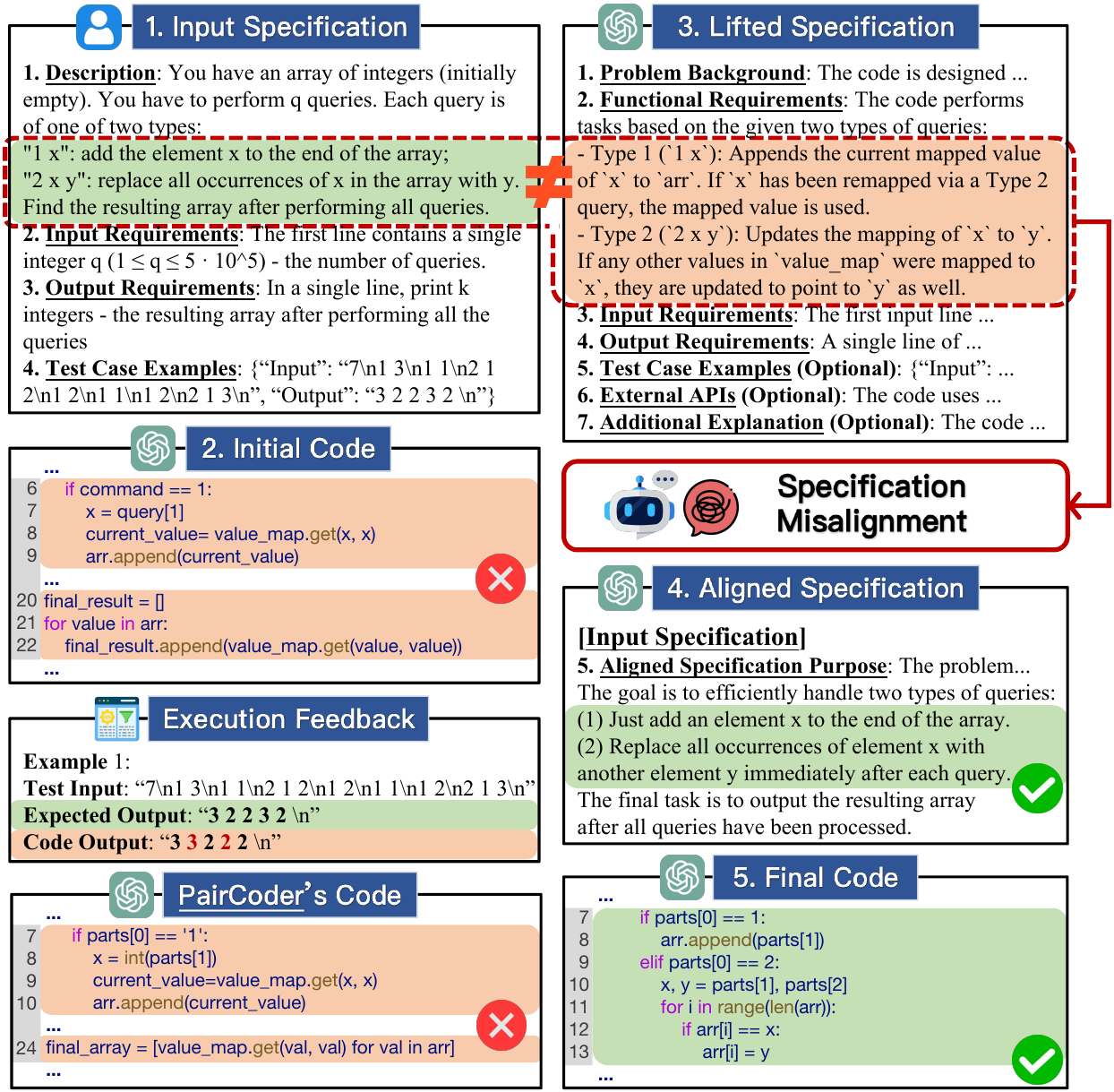}
    \vspace{-5mm}
    \caption{An example from Codeforces with GPT-4o-mini}
    \label{fig:motivation}
    \vspace{-4mm}
\end{figure}

To illustrate the key idea of specification alignment, we present a real-world example to analyze the necessity of two key components in \tech{}: specification lifting (Section~\ref{subsec:lifting}) and specification alignment (Section~\ref{subsec:alignment}).
The necessity of identifying misaligned specifications has been discussed in Section~\ref{sec:introduction}.

Figure~\ref{fig:motivation} presents a real-world example from the Codeforces~\cite{codeforces2023} platform. 
In this example, we employ an advanced LLM (GPT-4o-mini~\cite{openai2024introducing}) with Zero-shot learning~\cite{chen2021evaluating} to generate the initial code based on the input specification. 
However, due to the complexity of this problem, the generated code is incorrect.
Specifically, the LLM fails to accurately perceive the ``two query types'' in the specification, leading to the corresponding error in the initial code.
To further investigate this issue, we applied the state-of-the-art PairCoder~\cite{zhang2024paircoder} to adjust diverse solution plans and repair the code using execution feedback (shown in Figure~\ref{fig:motivation}).
However, PairCoder's code (shown in Figure~\ref{fig:motivation}) still exhibits the logical error.
In contrast, by aligning the purpose behind the ``two query types'' (shown in the aligned specification of Figure~\ref{fig:motivation}), the LLM successfully generates a correct implementation (i.e., final code in Figure~\ref{fig:motivation}). 
\textit{This motivates the potential of improving LLM-based code generation through specification alignment.}

Moreover, our natural attempt to directly instruct the LLM to align the specification is unsuccessful, still resulting in incorrect code. 
In contrast, explicitly extracting the LLM-perceived specification (i.e., lifted specification in Figure~\ref{fig:motivation}) enables the subsequent generation of correct code. 
This is because \tech{} systematically compares the lifted specification with the input specification, effectively analyzing specification misalignment for facilitating success alignment (i.e., the aligned specification in Figure~\ref{fig:motivation}). 
\textit{This motivates the necessity of explicitly obtaining the LLM-perceived specification.}

Although the specification lifting component effectively helps analyze specification misalignment, mitigating this misalignment remains a significant challenge. 
To explore potential solutions, we further instruct the LLM to generate ten candidate aligned specifications based on the lifted specification (as evaluated in the \techWoAR{} variant in RQ3).
However, none of the generated specifications successfully address the misalignment, and thus the generated code remains incorrect. 
\textit{This motivates the necessity of designing effective alignment rules for specification alignment}, which can finally enhance LLM-based code generation.
\section{Approach}
\label{sec:approach}

\begin{figure*}[t!]
    \centering
    \includegraphics[width=1.0\linewidth]{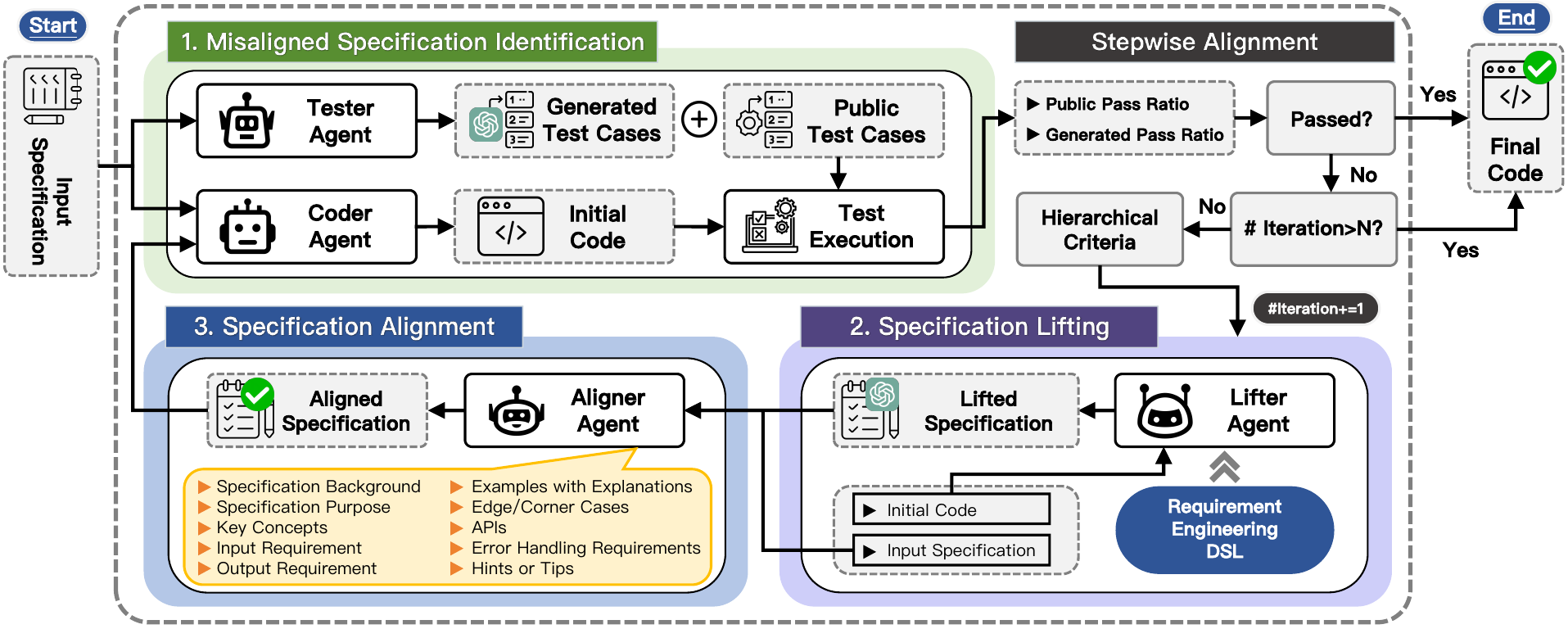}
    \vspace{-5mm}
    \caption{Overview of \tech{}}
    \label{fig:overview}
    \vspace{-4mm}
\end{figure*}

In this paper, we present \textbf{\tech{}}, a novel specification alignment technique designed to enhance the code generation performance of LLMs. 
Specifically, \tech{} automatically identifies misaligned specifications, lifts LLM-perceived specifications, and aligns them to generate correct code.
Figure~\ref{fig:overview} provides an overview of \tech{}, consisting of three main components: 
(1) \textbf{Misaligned Specification Identification} (Section~\ref{subsec:identification}) employs a coder agent to generate initial code and a tester agent to estimate its correctness to determine whether the LLM has correctly perceived the input specification.
(2) \textbf{Specification Lifting} (Section~\ref{subsec:lifting}) employs a novel LLM-driven lifting strategy to explicitly extract high-level perceived specifications from low-level generated code based on a pre-defined requirement DSL.
(3) \textbf{Specification Alignment} (Section~\ref{subsec:alignment}) applies ten pre-defined alignment rules to systematically align different ingredients of the misaligned specification, facilitating the subsequent code generation.
In the following, we provide a detailed description of each component in \tech{}.
Here, we reuse the real-world example introduced in Section~\ref{sec:motivating_example} for illustration.

\vspace{-2mm}
\subsection{Misaligned Specification Identification}
\label{subsec:identification}
In \tech{}, identifying misaligned specifications is a critical step that evaluates whether the LLM correctly perceives the input specification.
It is important to emphasize that the absence of this component would cause \tech{} to indiscriminately perform specification alignment on all input specifications. 
This indiscriminate processing could introduce significant drawbacks. 
First, for correctly LLM-perceived input specifications, forced alignment may lead to erroneous modifications due to the inherent hallucination issues of LLMs, potentially transforming originally correct code into incorrect implementations. 
Second, such redundant alignment operations impose unnecessary computational overhead, significantly increasing both time and token consumption.

\begin{figure}[t!]
    \centering
    \includegraphics[width=1.0\linewidth]{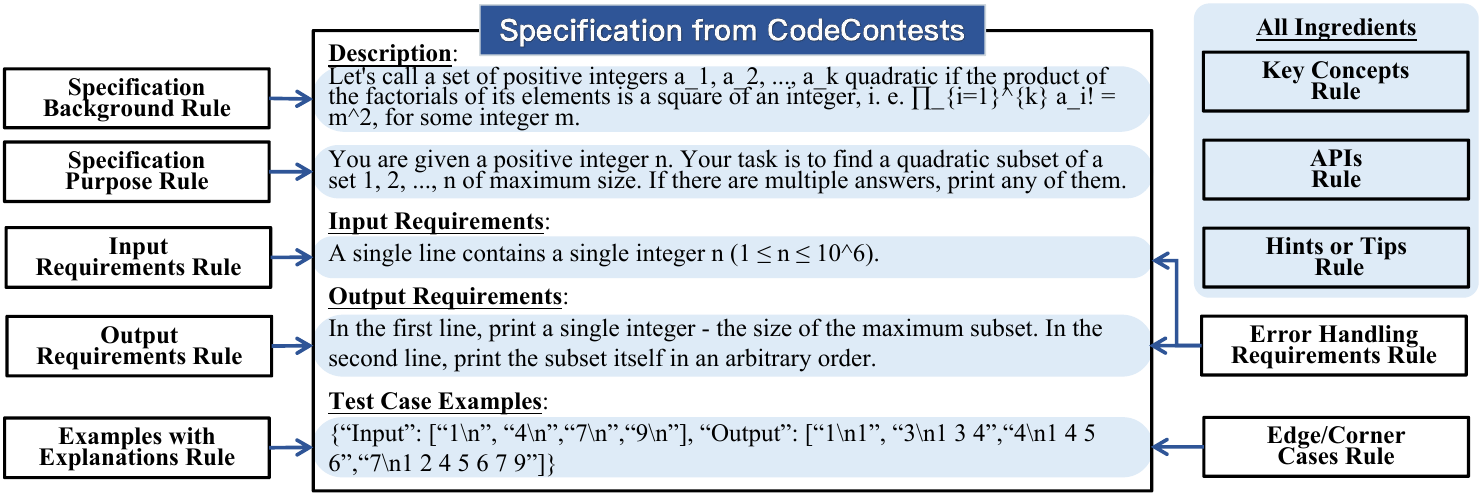}
    \vspace{-5mm}
    \caption{Correspondence between alignment rules and key specification ingredients}
    \label{fig:alignment_rules}
    \vspace{-4mm}
\end{figure}

As shown in Figure~\ref{fig:overview}, we design a dual-agent framework comprising a coder agent to generate the initial code (shown in Figure~\ref{fig:motivation}) and a tester agent to generate test cases. 
The correctness of the generated code serves as an indicator of the LLM's perception of the input specification~\cite{wei2024requirements,guo2025intention}.
In particular, the tester agent independently generates test cases based on the input specification, without relying on the initial code. 
This independence ensures the objectivity of test cases generation, mitigating any potential bias introduced by erroneous initial code. 
Following the existing research~\cite{huang2023agentcoder}, we carefully design a prompt to guide the tester agent in generating effective test cases. 
The prompt details are as follows:
\begin{figure}[H]
    \vspace{-3mm}
    \centering
    \includegraphics[width=1.0\linewidth]{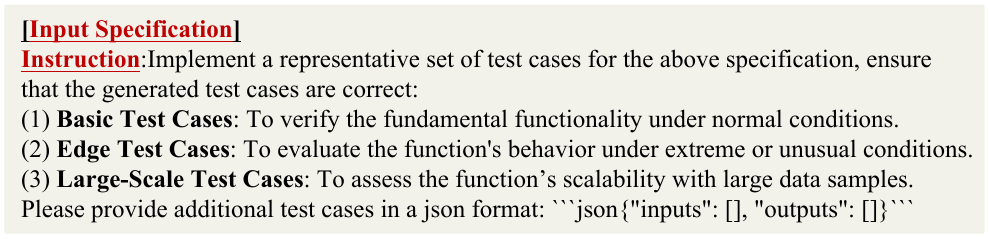}
    % \caption{xxx}
    \label{fig:tester_prompt}
    \vspace{-8mm}
\end{figure}

In practice, programming specifications commonly include public test cases. 
To comprehensively assess the generated code, \tech{} combines these public test cases with the additional test cases generated by the tester agent. 
These test cases can help ensure the reliability of this step for identifying misaligned specification. 
In Section~\ref{subsec:RQ3}, we further explore the influence of public test cases and generated test cases on the effectiveness of \tech{}, respectively.
Upon executing these test cases, \tech{} estimates the correctness of the initial code. 
If the code fails to pass all test cases, the input specification is deemed misaligned, activating the subsequent components for specification alignment. 
Conversely, if the code successfully passes all test cases, it is considered the final output of \tech{}.
We acknowledge that the LLM-generated test cases may contain partial inaccuracies and cannot guarantee complete correctness. 
Nevertheless, consistent with findings of prior studies~\cite{chen2022codet,lin2025flowgen}, they still contribute to enhancing the overall performance of LLM code generation. 
A systematic discussion of this phenomenon will be presented in Section~\ref{subsec:test_quality}.

\subsection{Specification Lifting}
\label{subsec:lifting}
After identifying misaligned input specifications based on the estimated correctness of the generated code, it is essential to thoroughly analyze the detailed misalignment by comparing the generated code and the input specification, which can facilitate specification alignment (as discussed in Section~\ref{sec:motivating_example}). 
However, directly comparing and analyzing the specification misalignment is inherently challenging due to the significant difference in abstraction levels between the initial code (i.e., low-level programming language) and the input specification (i.e., high-level requirement description). 
To effectively bridge this semantic gap, we design a specification lifting component that explicitly extracts the LLM-perceived specification from the generated code.

Inspired by existing lifting research~\cite{altinay2020binrec,naus2024poster,zhang2025towards}, we propose a novel lifter agent (illustrated in Figure~\ref{fig:overview}) that lifts the generated code into a DSL tailored for requirement specifications. 
Compared to ordinary natural language descriptions, a requirement DSL provides a standardized representation of critical software requirements, mitigating common issues such as ambiguity and incompleteness. 
Building on research in software requirements engineering~\cite{altinay2020binrec,naus2024poster,zhang2025towards}, we define a tailored requirement DSL for our task, encompassing the following key attributes:
\begin{figure}[H]
    \vspace{-3mm}
    \centering
    \includegraphics[width=1.0\linewidth]{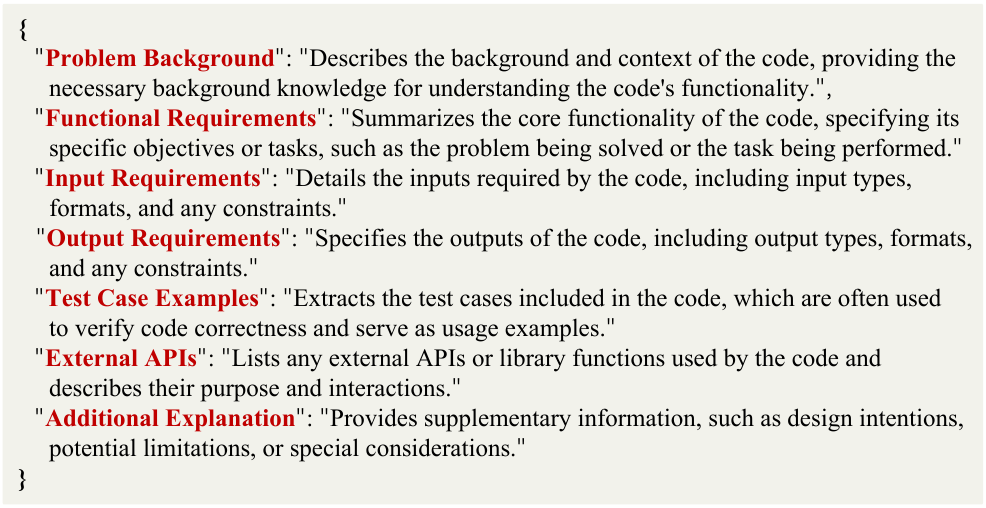}
    \label{fig:tester_prompt}
    \vspace{-8mm}
\end{figure}
\noindent
This highly standardized requirement DSL aligns with \textit{IEEE specification standard}~\cite{doe2011recommended}, ensuring that the extracted lifted specification(shown in Figure~\ref{fig:motivation}) comprehensively and effectively represents the LLM-perceived specification from the initial code.
\revision{Furthermore, we discuss the scalability of the DSL design in Section~\ref{subsec:dsl_scalability}, highlighting its potential to further enhance the generalizability and applicability of our approach.}

Moreover, we design a prompt for the lifter agent that consists of the initial code, the definition of requirement DSL, and a task instruction (``Analyze the initial code exclusively and translate it as the lifted specification based on the defined requirement DSL.'').
Through LLM-driven specification lifting, the lifter agent generates the requirement DSL and parses it into a structured lifted specification template, thereby producing the final lifted specification. 
Particularly, our ablation results (in Section~\ref{subsec:RQ3}) demonstrate that the specification lifting strategy significantly outperforms the commonly-used test execution feedback strategy in existing agent-based code generation techniques.

\subsection{Specification Alignment}
\label{subsec:alignment}
The core idea of this component is to analyze potential misaligned ingredients between the LLM-perceived specification (i.e., the lifted specification) and the input specification, and then mitigate these misalignment using pre-defined alignment rules. 
Based on the existing studies in software requirements engineering~\cite{greenspan1994formal,glinz2000problems,doe2011recommended}, we define ten alignment rules that systematically align different kinds of ingredients in the misaligned specification (shown in Figure~\ref{fig:alignment_rules}).
These alignment rules are detailed as follows:

\begin{itemize}[leftmargin=10pt]
    \item \textbf{Specification Background}: further explains the background, motivation, or domain-specific knowledge required by the specification. 
    This helps the LLM fully perceive the problem context for more accurate code generation.
    
    \item \textbf{Specification Purpose}: emphasizes the detailed objective or core tasks of the specification. 
    This helps the LLM maintain focus on the intended goal, reducing the likelihood of generating code that deviates from the required functionality.
    
    \item \textbf{Key Concepts}: identifies and explains the critical terminology in the specification, ensuring that the LLM accurately perceives the core terms and concepts, thereby minimizing code errors caused by conceptual misperception.
    
    \item \textbf{Input Requirements}: emphasizes the input requirements of the specification, including data types, formats, and constraints (e.g., value ranges, size limits, or specific conditions). 
    It helps the generated code correctly process input data.
    
    \item \textbf{Output Requirements}: emphasizes the output requirements of the specification, including data types, formats, and constraints (e.g., precision, delimiters, or sorting rules).
    It helps the generated code correctly produce the expected output.
    
    \item \textbf{Examples with Explanations}: provides step-by-step analysis of test cases, illustrating the detailed logic from input to output. 
    This enhances the LLM's perception of the programming logic.
    
    \item \textbf{Edge/Corner Cases}: introduces three additional LLM-generated test cases covering extreme or boundary conditions to ensure that the generated code handles unusual edge cases correctly.
    
    \item \textbf{APIs}: specifies external APIs or library functions relevant to the task, including their names and functionalities. 
    This helps the LLM implement the required functions more effectively.
    
    \item \textbf{Error Handling Requirements}: defines the expected behavior when encountering invalid inputs, such as returning default values, raising exceptions, or applying special mechanisms. 
    It ensures the code behaves as expected in exceptional circumstances.
    
    \item \textbf{Hints or Tips}: provides optional implementation suggestions, such as recommended algorithms or data structures. 
    It serves as a supplementary guideline for addressing any remaining specification misalignment not covered by the preceding nine rules.
\end{itemize}

Furthermore, \tech{} employs an LLM-based aligner agent to facilitate specification alignment. 
Specifically, we design a prompt for the aligner agent, which consists of the input specification, the lifted specification, the ten pre-defined alignment rules, and a task instruction (``Analyze the input specification and the lifted specification to identify misalignment or omissions. Select one or more ingredients from the ten alignment rules to improve the input specification.'').
The aligner agent automatically generates the aligned ingredients for the input specification, which collectively constitute the aligned specification (as shown in Figure~\ref{fig:motivation}).

In particular, the specification alignment is implemented as a stepwise search process. 
In each iteration, \tech{} produces an aligned specification and subsequently uses it to generate new code.
To measure the quality of different versions of generated code, we design a hierarchical criteria based on two pass ratios: 
(1) the primary pass ratio, which measures the pass ratio of public test cases provided in the input specification, and (2) the secondary pass ratio, which measures the pass ratio of LLM-generated test cases.
When comparing different code, the primary pass ratio is considered first; only if this ratio is identical is the secondary pass ratio considered.
This hierarchical criterion prioritizes the primary pass ratio, as the correctness of public test cases is guaranteed, whereas LLM-generated test cases may contain errors. 
Importantly, if the new code generated from the current aligned specification achieves a higher hierarchical pass ratio than the previous iteration (even if it does not fully pass all test cases), \tech{} retains the current aligned ingredients and proceeds to the next iteration. 
This greedy optimization process ensures stepwise improvements during specification alignment process. 
In the future, more sophisticated search algorithms could be integrated to further enhance this component.

Besides, \tech{} terminates the alignment process once a pre-defined number of iterations (denoted as $N$) is reached, outputting the code with the highest hierarchical pass ratio. 
The number of iterations is a critical hyper-parameter for all agent-based techniques, and its impact will be discussed in Section~\ref{subsec:RQ2}.
Through this specification alignment component, \tech{} can effectively achieve specification alignment for generating correct code.

\section{Evaluation Design}
\label{sec:evaluation_design}
Our study aims to addresses the following research questions (RQs):
\begin{itemize}[leftmargin=10pt]
\item \textbf{RQ1}: How does \tech{} perform in terms of effectiveness and efficiency compared to the state-of-the-art techniques?
\item \textbf{RQ2}: How do hyper-parameters affect \tech{}'s effectiveness?
\item \textbf{RQ3}: How does each main component in \tech{} contribute to the overall effectiveness?
\end{itemize}

\subsection{Benchmarks}
\label{subsec:benchmarks}
To comprehensively evaluate \tech{}, we utilize five challenging benchmarks in our study: APPS~\cite{hendrycks2021measuring}, APPS-Eval~\cite{dong2024codescore}, CodeContests-Raw~\cite{li2022competition}, CodeContests~\cite{li2022competition}, and xCodeEval~\cite{khan2023xcodeeval}. 
These datasets are commonly used in many existing LLM-based code generation studies~\cite{jain2023improving,islam2024mapcoder,tian2025fixing}.
Notably, we do not include basic programming datasets such as HumanEval~\cite{chen2021evaluating} and MBPP~\cite{austin2021program}, as LLMs already achieve a Pass@1 exceeding 95\% on these datasets with simple Zero-shot~\cite{evalplusleaderboard2024} setting.
Instead, we select more challenging competition-level datasets to better assess the performance of \tech{}.

\textit{\underline{APPS}} consists of programming problems collected from various competitive programming platforms (e.g., LeetCode~\cite{leetcode2023}), including 5,000 training data and 5,000 test data.
To balance the evaluation cost and conclusion generalizability, we randomly sample 300 problems from test set based on the difficulty distribution, following the existing work~\cite{olausson2024selfrepair,tian2025fixing}. 
Additionally, to improve the comprehensiveness of the evaluation, APPS is extended into \textit{\underline{APPS-Eval}} by constructing over 100 additional test cases for each problem.

\textit{\underline{CodeContests-Raw}} is proposed by Google DeepMind~\cite{googledeepmind2025}, consisting of 13,328 training data, 117 validation data, and 165 test data.
It is designed to evaluate the ability of LLMs to solve challenging programming problems. 
We also use the extended version, \textit{\underline{CodeContests}}, which includes $\sim$190 additional test cases over the raw version. 
Following existing studies~\cite{chen2022codet,zhangplanning}, we use all 165 test data from both CodeContests-Raw and CodeContests.

\textit{\underline{xCodeEval}} is a competition-level multi-task benchmark, consisting of $\sim$7,500 programming problems and 25 million code examples collected from Codeforces~\cite{codeforces2023}. 
To maintain consistency with APPS, we randomly sample 300 problems from the code generation test data based on the frequency distribution of difficulty levels.

These datasets provide several public test cases (as part of the original specification), along with separate private test cases used for evaluating code correctness. 
Some studied baselines (including Self-repair~\cite{olausson2024selfrepair}, $\mu$FiX~\cite{tian2025fixing}, Self-collaboration~\cite{dong2024selfcollaboration}, PairCoder~\cite{zhang2024paircoder}, and TGen~\cite{mathews2024tgen}) rely on the execution result of public test cases. 
However, in APPS, public test cases are presented in natural language within the original specification, rendering them non-executable. 
Hence, for APPS, we randomly sample three test cases from private test cases to serve as executable public test cases.
To prevent result inflation due to potential test case leakage, we remove the three cases from private test cases for performance evaluation.

% \vspace{-2mm}
\subsection{Metrics}
\label{subsec:metrics}
Following existing studies~\cite{jiang2024self,tian2025fixing}, we use \textit{\underline{Pass@$k$}} and \textit{\underline{AvgPassRatio}} to evaluate the effectiveness of \tech{}. 
Pass@$k$ measures the functional correctness of the generated code. 
For a given programming problem, the LLM generates $k$ code instances. 
If any of the instances passes all the private test cases, the problem is considered solved. 
Pass@$k$ is the percentage of problems solved out of the total number of problems. 
As demonstrated by existing studies~\cite{dong2024selfcollaboration,mu2024clarifygpt,chen2024teaching}, developers tend to consider the first code instance generated by the LLM, and thus we set $k$ = 1 following existing work~\cite{zhang2024paircoder,tian2025fixing,lin2025flowgen}. 
Note that Pass@1 is a stricter criterion, making improvements in this metric both challenging and practically meaningful.
AvgPassRatio measures the degree of correctness of the generated code on private test cases, and differs from Pass@$k$ that evaluates whether the generated code is completely correct on private test cases. 
AvgPassRatio calculates the ratio of passing private test cases for each problem, and then computes the average ratio across all problems. 
Both metrics are largely complementary, higher Pass@$k$ and AvgPassRatio values indicate better effectiveness.

Additionally, we evaluate the efficiency of \tech{} by measuring both \textit{\underline{time overhead}} and \textit{\underline{token overhead}} (which includes prompt token cost and generated token cost) following existing work~\cite{hong2023metagpt,dong2024selfcollaboration,tian2025fixing}.
Smaller values of time and token overhead indicate better efficiency in code generation.

% \vspace{-2mm}
\subsection{Compared Techniques}
\label{subsec:baselines}
To thoroughly evaluate \tech{}, we compare it with six representative or state-of-the-art agent-based techniques: \textbf{Self-collaboration}~\cite{dong2024selfcollaboration}, \textbf{MetaGPT}~\cite{hong2023metagpt}, \textbf{AgentCoder}~\cite{huang2023agentcoder}, \textbf{TGen}~\cite{mathews2024tgen}, \textbf{FlowGen}~\cite{lin2025flowgen}, and \textbf{PairCoder}~\cite{zhang2024paircoder}. 
They share the same high-level insight, which simulates the software development process by designing diverse agents. 
The primary difference among these techniques lies in the roles assigned to agents and the distinct strategies to address corresponding sub-tasks.
Although our study focuses on agent-based techniques, for a more comprehensive evaluation of \tech{}, we also compare it with four typical or state-of-the-art prompt-based code generation techniques, including \textbf{Zero-shot}~\cite{chen2021evaluating}, \textbf{SCoT}~\cite{li2025structured}, \textbf{Self-repair}~\cite{olausson2024selfrepair}, and \textbf{$\mu$FiX}~\cite{tian2025fixing}. 
Due to space limitations, more details on these techniques can be found in their respective papers.

% \vspace{-2mm}
\subsection{Implementation Details}
\label{subsec:implementation}
To evaluate the performance of \tech{}, we select a variety of open-source and commercial LLMs for our study.
For open-source LLMs, we chose two representative LLMs: Qwen-Coder~\cite{hui2024qwen2} (version \textit{qwen2.5-coder-7b-instruct}) and DeepSeek-Coder~\cite{guo2024deepseek} (version \textit{deepseek-coder-7b-instruct-v1.5}). 
Specifically, we download them via the Huggingface~\cite{wolf2019huggingface} platform and deploy them in a local environment for our experiments.
For commercial LLMs, we select two widely-used and advanced LLMs: GPT-4o~\cite{openai2024introducing} (version \textit{gpt-4o-mini-2024-07-18}) and Gemini-1.5~\cite{team2024gemini} (version \textit{gemini-1.5-flash-002}). 
We access these commercial LLMs through the respective APIs provided by OpenAI~\cite{openai2025} and Google AI~\cite{googleai2025}.
These LLMs, which have demonstrated excellent performance in code generation and are widely adopted~\cite{zhang2024paircoder,gao2024search,tian2024large}, evaluate the generalizability of \tech{} to some extent.
In our experiments, we set max tokens to 1024 and temperature to 0.8 for all LLMs. 
The maximum number of iterations $N$ is set to 10 for \tech{} and baselines.
As LLMs often generate code that includes natural language text segments (e.g., explanatory text), which can lead to compilation failures or execution errors.
Following existing studies~\cite{liu2023your,tian2025fixing}, we utilize a code sanitizer tool~\cite{codesanitizer2025} to preprocess the code generated by the LLMs. 
It can automatically detect and remove non-functional text segments from the generated code.

\section{Results and Analysis}
\label{sec:results_and_analysis}

\begin{table*}[t!]
    \caption{Effectiveness and efficiency comparison in terms of Pass@1 ($\uparrow$), AvgPassRatio ($\uparrow$), Time Overhead ($\downarrow$), and Token Overhead ($\downarrow$). APR is short for AvgPassRatio.}
    \vspace{-3mm}
    \label{tab:rq1}
    \centering
    \tabcolsep=0.80mm
    % \normalsize
    \small
    \begin{adjustbox}{max width=1.0 \textwidth,center}
        \begin{tabular}{llcccccccccccc}
            % \toprule
            \toprule
        	\multirow{2}{*}{\textbf{LLM}} & \multirow{2}{*}{\textbf{Technique}} & \multicolumn{2}{c}{\textbf{APPS}} & \multicolumn{2}{c}{\textbf{APPS-Eval}} & \multicolumn{2}{c}{\textbf{CodeContests-Raw}} &
            \multicolumn{2}{c}{\textbf{CodeContests}} & \multicolumn{2}{c}{\textbf{xCodeEval}} & \multirow{2}{*}{\textbf{\makecell{Time \\ (h)}}} & \multirow{2}{*}{\textbf{\makecell{Token \\ (M)}}} \\ \cmidrule(lr){3-4} \cmidrule(lr){5-6} \cmidrule(lr){7-8} \cmidrule(lr){9-10} \cmidrule(lr){11-12}
        	& & \textbf{Pass@1} & \textbf{APR} & \textbf{Pass@1} & \textbf{APR} & \textbf{Pass@1} & \textbf{APR} & \textbf{Pass@1} & \textbf{APR} & \textbf{Pass@1} & \textbf{APR} & & \\
        	\midrule
            
            \multirow{11}{*}{DeepSeek-Coder-v1.5} & Zero-shot & 18.00\% & 34.25\% & 8.33\% & 33.35\% & 7.27\% & 14.78\% & 2.42\% & 13.02\% & 11.00\% & 24.77\% & 3.50  & 0.64 \\
            & SCoT & 19.00\% & 35.95\% & 8.67\% & 34.82\% & 7.27\% & 16.75\% & 3.64\% & 14.25\% & 12.33\% & 25.21\% & 5.15  & 2.43  \\
            & Self-repair & 20.67\% & 38.38\% & 8.67\% & 36.23\% & 7.27\% & 17.33\% & 3.64\% & 14.73\% & 12.67\% & 26.65\% & 6.39  & 2.06  \\
            & $\mu$FiX & 23.00\% & 40.42\% & 10.67\% & 39.04\% & 9.70\% & 19.90\% & 4.24\% & 15.10\% & 13.33\% & 27.12\% & 18.94  & 6.29  \\ \cdashline{2-14}
            & Self-collaboration & 23.00\% & 35.37\% & 10.33\% & 32.84\% & 9.70\% & 18.48\% & 4.24\% & 15.96\% & 13.33\% & 24.83\% & 87.21  & 18.48  \\
            & MetaGPT & 24.33\% & 38.27\% & 10.67\% & 34.84\% & 7.27\% & 13.79\% & 4.85\% & 12.52\% & 12.67\% & 20.88\% & 127.43  & 25.16  \\
            & AgentCoder & 21.67\% & 37.53\% & 9.33\% & 33.59\% & 9.09\% & 16.01\% & 4.24\% & 14.74\% & 12.00\% & 25.34\% & 108.80  & 15.67  \\
            & TGen & 22.33\% & 36.34\% & 10.00\% & 33.37\% & 7.88\% & 17.41\% & 3.03\% & 13.79\% & 15.00\% & 27.88\% & 73.44  & 17.36  \\
            & FlowGen & 23.67\% & 41.09\% & 10.67\% & 38.86\% & 9.70\% & 15.87\% & 4.24\% & 14.36\% & 15.33\% & 29.30\% & 110.95  & 20.43  \\
            & PairCoder & 23.67\% & 37.37\% & 9.67\% & 35.09\% & 10.30\% & 20.82\% & 4.24\% & 18.22\% & 13.33\% & 26.64\% & 75.99  & 16.93  \\
            & \textbf{\tech{}} & \textbf{35.33\%} & \textbf{59.62\%} & \textbf{14.33\%} & \textbf{50.99\%} & \textbf{13.33\%} & \textbf{28.67\%} & \textbf{7.88\%} & \textbf{30.16\%} & \textbf{20.67\%} & \textbf{45.58\%} & 60.22  & 15.16  \\ \midrule
            
            \multirow{11}{*}{Qwen2.5-Coder} & Zero-shot & 19.67\% & 36.66\% & 9.33\% & 34.95\% & 7.88\% & 17.97\% & 5.45\% & 18.64\% & 10.67\% & 27.63\% & 3.23  & 0.54 \\
            & SCoT & 21.67\% & 38.29\% & 10.00\% & 35.94\% & 8.48\% & 15.67\% & 6.06\% & 15.13\% & 11.33\% & 22.52\% & 6.62  & 2.37  \\
            & Self-repair & 21.67\% & 37.12\% & 10.00\% & 36.05\% & 8.48\% & 19.82\% & 6.06\% & 19.93\% & 12.00\% & 24.17\% & 5.54  & 1.54 \\
            & $\mu$FiX & 24.33\% & 39.34\% & 11.00\% & 41.43\% & 10.91\% & 22.57\% & 6.67\% & 20.97\% & 13.33\% & 27.92\% & 19.99  & 6.08  \\ \cdashline{2-14}
            & Self-collaboration & 24.00\% & 35.88\% & 10.67\% & 33.24\% & 9.09\% & 20.82\% & 7.27\% & 22.21\% & 12.33\% & 27.08\% & 62.60  & 15.75  \\
            & MetaGPT & 23.33\% & 36.04\% & 11.00\% & 33.78\% & 9.70\% & 18.17\% & 7.27\% & 19.38\% & 11.33\% & 23.92\% & 110.45  & 25.54  \\
            & AgentCoder & 24.33\% & 38.06\% & 11.33\% & 33.76\% & 12.12\% & 22.36\% & 9.09\% & 22.52\% & 14.33\% & 28.27\% & 139.57  & 16.90  \\
            & TGen & 26.33\% & 37.10\% & 11.00\% & 31.75\% & 9.09\% & 13.62\% & 6.06\% & 12.11\% & 14.33\% & 22.21\% & 74.42  & 17.11  \\
            & FlowGen & 25.33\% & 38.61\% & 10.00\% & 34.60\% & 8.48\% & 15.60\% & 6.67\% & 14.87\% & 12.00\% & 24.72\% & 108.02  & 21.98  \\
            & PairCoder & 30.33\% & 50.20\% & 11.67\% & 42.18\% & 13.94\% & 29.07\% & 9.09\% & 25.53\% & 16.00\% & 37.33\% & 84.81  & 15.64  \\
            & \textbf{\tech{}} & \textbf{37.67\%} & \textbf{60.63\%} & \textbf{16.67\%} & \textbf{51.53\%} & \textbf{15.15\%} & \textbf{32.44\%} & \textbf{10.30\%} & \textbf{34.11\%} & \textbf{18.67\%} & \textbf{42.09\%} & 51.20  & 14.38  \\ \midrule

            \multirow{11}{*}{GPT-4o-mini} & Zero-shot & 33.67\% & 43.61\% & 13.33\% & 41.38\% & 8.48\% & 18.16\% & 4.85\% & 17.96\% & 23.67\% & 32.82\% & 1.67  & 0.61 \\
            & SCoT & 35.33\% & 46.52\% & 14.33\% & 43.03\% & 11.52\% & 23.73\% & 6.06\% & 22.00\% & 25.00\% & 33.68\% & 1.43  & 2.46  \\
            & Self-repair & 36.67\% & 47.01\% & 14.67\% & 43.24\% & 10.30\% & 19.94\% & 6.67\% & 20.66\% & 25.67\% & 34.88\% & 1.97  & 1.63  \\
            & $\mu$FiX & 40.33\% & 49.31\% & 15.67\% & 46.15\% & 12.12\% & 25.53\% & 9.09\% & 23.21\% & 28.67\% & 38.95\% & 5.62  & 6.55  \\ \cdashline{2-14}
            & Self-collaboration & 42.00\% & 55.19\% & 16.33\% & 48.25\% & 14.55\% & 27.04\% & 10.91\% & 25.72\% & 30.67\% & 45.48\% & 31.40  & 17.00  \\
            & MetaGPT & 41.33\% & 49.90\% & 16.33\% & 45.83\% & 12.12\% & 25.69\% & 9.70\% & 25.78\% & 26.00\% & 35.14\% & 32.90  & 27.20  \\
            & AgentCoder & 41.33\% & 49.16\% & 16.67\% & 44.21\% & 15.76\% & 26.58\% & 10.30\% & 26.48\% & 26.00\% & 33.46\% & 28.71  & 16.84  \\
            & TGen & 41.00\% & 47.16\% & 16.67\% & 42.62\% & 9.09\% & 13.17\% & 7.88\% & 12.82\% & 30.67\% & 37.58\% & 48.21  & 18.32  \\
            & FlowGen & 43.67\% & 52.32\% & 17.00\% & 45.43\% & 12.12\% & 21.43\% & 9.09\% & 19.35\% & 30.00\% & 38.00\% & 32.70  & 21.33  \\
            & PairCoder & 49.00\% & 63.68\% & 19.67\% & 56.02\% & 18.79\% & 34.14\% & 11.52\% & 31.88\% & 31.67\% & 46.06\% & 29.87  & 16.71  \\
            & \textbf{\tech{}} & \textbf{56.33\%} & \textbf{72.98\%} & \textbf{22.67\%} & \textbf{63.55\%} & \textbf{23.64\%} & \textbf{40.65\%} & \textbf{17.58\%} & \textbf{42.10\%} & \textbf{38.67\%} & \textbf{57.34\%} & 24.62  & 14.41  \\ \midrule
            
            \multirow{11}{*}{Gemini-1.5-Flash} & Zero-shot & 47.67\% & 59.33\% & 19.67\% & 51.65\% & 29.09\% & 39.30\% & 23.64\% & 33.49\% & 27.00\% & 41.44\% & 1.10  & 0.50  \\
            & SCoT & 48.67\% & 59.51\% & 20.67\% & 51.10\% & 30.30\% & 40.08\% & 25.45\% & 35.05\% & 28.00\% & 41.26\% & 1.13  & 2.22  \\
            & Self-repair & 49.33\% & 60.14\% & 21.33\% & 52.71\% & 32.33\% & 42.64\% & 25.45\% & 36.71\% & 29.33\% & 42.72\% & 1.50  & 0.99  \\
            & $\mu$FiX & 51.33\% & 62.05\% & 22.33\% & 56.69\% & 34.39\% & 44.22\% & 28.48\% & 40.93\% & 31.00\% & 44.10\% & 4.08  & 5.10  \\ \cdashline{2-14}
            & Self-collaboration & 54.33\% & 66.28\% & 22.67\% & 54.89\% & 36.97\% & 49.11\% & 30.30\% & 43.13\% & 31.67\% & 46.19\% & 21.74  & 10.71  \\
            & MetaGPT & 55.00\% & 65.31\% & 22.67\% & 53.60\% & 35.76\% & 45.68\% & 28.48\% & 40.34\% & 29.00\% & 35.21\% & 21.74  & 22.79  \\
            & AgentCoder & 53.33\% & 60.51\% & 22.33\% & 51.23\% & 42.42\% & 50.90\% & 32.73\% & 47.17\% & 30.67\% & 37.55\% & 20.22  & 13.83  \\
            & TGen & 55.67\% & 68.96\% & 23.33\% & 58.76\% & 36.97\% & 50.06\% & 26.67\% & 42.82\% & 34.00\% & 47.12\% & 42.36  & 10.43 \\
            & FlowGen & 53.33\% & 63.15\% & 21.67\% & 51.01\% & 40.00\% & 54.31\% & 31.52\% & 48.92\% & 29.67\% & 42.14\% & 22.22  & 13.89  \\
            & PairCoder & 55.00\% & 68.47\% & 23.00\% & 56.82\% & 43.03\% & 54.42\% & 30.91\% & 47.05\% & 32.33\% & 46.72\% & 21.70  & 10.46  \\
            & \textbf{\tech{}} & \textbf{65.33\%} & \textbf{78.15\%} & \textbf{29.33\%} & \textbf{65.91\%} & \textbf{44.85\%} & \textbf{56.96\%} & \textbf{36.36\%} & \textbf{54.34\%} & \textbf{38.67\%} & \textbf{58.42\%} & 17.61  & 9.77  \\ 
            % \midrule
            \bottomrule
        \end{tabular}
    \end{adjustbox}
    \vspace{-4mm}
\end{table*}

\subsection{RQ1: Effectiveness and Efficiency}
\label{subsec:RQ1}

\subsubsection{Process:}
To answer RQ1, we apply \tech{} and 10 compared techniques to four studied LLMs.
We then evaluate the effectiveness of each technique across five widely-used benchmarks in terms of Pass@1 and AvgPassRatio.
In addition, we assess the efficiency of each technique in terms of time and token overhead.

\subsubsection{Results:}
Table~\ref{tab:rq1} shows the effectiveness and efficiency comparison results of these techniques. 
First, we observe that, on average, agent-based code generation techniques outperform prompt-based code generation techniques in terms of Pass@1 and AvgPassRatio.
This result confirms the effectiveness of the agent-based framework and further motivates the design of our \tech{}.

In particular, \textbf{\tech{} achieves the best effectiveness among all studied techniques, consistently demonstrating superior performance in both effectiveness metrics across all 20 subjects (4 LLMs $\times$ 5 benchmarks)}.
Specifically, \tech{} significantly improves all baselines by 29.60\%$\sim$93.55\% and 27.95\%$\sim$79.12\% in terms of Pass@1 and AvgPassRatio on average across all subjects, respectively.
Furthermore, the \textit{Wilcoxon Signed-Rank Test}~\cite{wilcoxon1970critical} at a significance level of 0.05 confirms that all p-values are smaller than \num{2.40e-7}, demonstrating the statistically significant superiority of \tech{} over all baselines in terms of Pass@1 and AvgPassRatio.

In addition, prompt-based code generation techniques outperform agent-based techniques in terms of both efficiency metrics (i.e., time and token overhead).
However, considering the significant effectiveness of agent-based techniques, a certain degree of additional cost is justified, illustrating their excellent balance of cost and effectiveness.
Notably, \textbf{our \tech{} achieves superior efficiency compared to all six agent-based baselines in terms of both efficiency metrics}.
Specifically, \tech{} improves all agent-based baselines by 22.44\%$\sim$39.14\% and 9.69\%$\sim$46.89\% in terms of time and token overhead on average across all subjects, respectively.
Furthermore, the \textit{Wilcoxon Signed-Rank Test}~\cite{wilcoxon1970critical} at a significance level of 0.05 confirms that all p-values are smaller than \num{1.56e-2}, demonstrating the statistically significant superiority of \tech{} over all agent-based baselines in terms of time and token overhead.

\begin{figure*}[t!]
    \centering
    \includegraphics[width=1.0\linewidth]{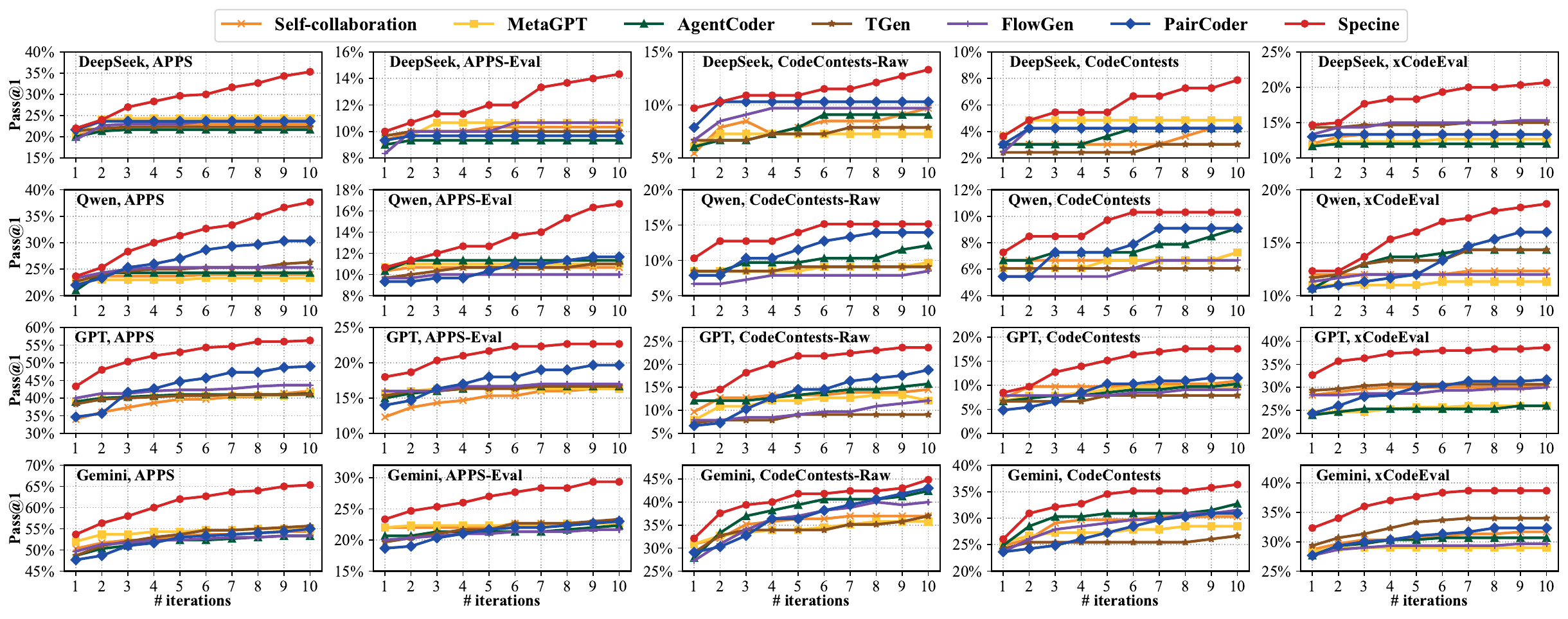}
    \vspace{-5mm}
    \caption{Influence of the number of iterations ($N$) in terms of Pass@1 ($\uparrow$)}
    \label{fig:RQ2-Pass@1}
    \vspace{-4mm}
\end{figure*}

\subsection{RQ2: Influence of Hyper-parameter}
\label{subsec:RQ2}

\subsubsection{Setup:}
The number of iterations is a key hyper-parameter in agent-based code generation techniques~\cite{huang2023agentcoder,zhang2024paircoder}.
In this research question, we investigate the influence of the number of iterations (i.e., $N$) on the effectiveness of \tech{}.
Specifically, for $1 \le N \le 10$, we evaluate the performance of \tech{} alongside six agent-based baselines across all 20 subjects (4 LLMs $\times$ 5 benchmarks) in terms of Pass@1 and AvgPassRatio.

\subsubsection{Results:}
Figure~\ref{fig:RQ2-Pass@1} illustrates the variation in Pass@1 across different agent-based techniques as the number of iterations increases.
Due to space limitations, we put the results for AvgPassRatio on our homepage~\cite{homepage2025}.
First, we observe that as the number of iterations increases, all agent-based techniques exhibit improvements in effectiveness for both Pass@1 and AvgPassRatio. 
However, across all iteration settings, \tech{} consistently outperforms all baselines, achieving 27.29\%$\sim$42.86\% higher Pass@1 and 25.86\%$\sim$68.22\% higher AvgPassRatio on average across all subjects.
This demonstrates the stable superiority of \tech{} across different iteration settings.
Furthermore, the \textit{Wilcoxon Signed-Rank Test}~\cite{wilcoxon1970critical} at a significance level of 0.05 confirms that all p-values are smaller than \num{1.09e-4}, demonstrating the statistically significant superiority of \tech{} under varying iteration settings.

In addition, the improvement achieved by increasing the number of iterations in \tech{} surpasses that of all baselines. 
Specifically, as $N$ increases from 1 to 10, \tech{} achieves an average improvement of 47.97\% in terms of Pass@1 across all subjects, whereas other baselines show an average improvement of 13.27\%$\sim$45.40\%. 
Similarly, for the AvgPassRatio metric, \tech{} demonstrates an average improvement of 37.02\%, while other baselines achieve an average improvement of 9.97\%$\sim$35.87\%.

In particular, these baselines appear to reach an effectiveness plateau after a certain number of iterations, such as when $N \ge 3$ on APPS with DeepSeek-Coder. 
In contrast, \tech{} consistently exhibits an upward trajectory across all iterations on almost all subjects, demonstrating its ability to leverage additional iterations for continuous improvement in effectiveness.
\revision{
This is because \tech{}'s specification lifting component keeps improving alignment over time. 
Even when early iterations do not achieve perfect alignment, \tech{} can extract detailed information from newly generated code to progressively refine the LLM-perceived specification.
In contrast, existing baselines rely primarily on coarse test feedback, which is less informative and may even lead to regression issues.}
We hypothesize that increasing the number of iterations may further enhance \tech{}'s performance; however, this entails a trade-off between effectiveness and computational cost.

\begin{table*}[t!]
    \caption{Comparison between \tech{} and its variants in terms of Pass@1 ($\uparrow$) and AvgPassRatio ($\uparrow$). APR is short for AvgPassRatio.}
    \vspace{-3mm}
    \label{tab:rq3}
    \centering
    \tabcolsep=1.80mm
    % \normalsize
    \small
    \begin{adjustbox}{max width=1.0 \textwidth,center}
        \begin{tabular}{llcccccccccc}
            % \toprule
            \toprule
        	\multirow{2}{*}{\textbf{LLM}} & \multirow{2}{*}{\textbf{Technique}} & \multicolumn{2}{c}{\textbf{APPS}} & \multicolumn{2}{c}{\textbf{APPS-Eval}} & \multicolumn{2}{c}{\textbf{CodeContests-Raw}} &
            \multicolumn{2}{c}{\textbf{CodeContests}} & \multicolumn{2}{c}{\textbf{xCodeEval}} \\ \cmidrule(lr){3-4} \cmidrule(lr){5-6} \cmidrule(lr){7-8} \cmidrule(lr){9-10} \cmidrule(lr){11-12}
        	& & \textbf{Pass@1} & \textbf{APR} & \textbf{Pass@1} & \textbf{APR} & \textbf{Pass@1} & \textbf{APR} & \textbf{Pass@1} & \textbf{APR} & \textbf{Pass@1} & \textbf{APR} \\
        	\midrule
            
            \multirow{5}{*}{DeepSeek-Coder-v1.5} & \techWoPTC{} & 29.67\% & 46.70\% & 12.00\% & 41.41\% & 11.52\% & 22.10\% & 6.06\% & 18.89\% & 17.33\% & 32.52\% \\
            & \techWoT{} & 32.00\% & 54.82\% & 13.33\% & 48.42\% & 11.52\% & 23.47\% & 6.06\% & 25.04\% & 16.00\% & 36.93\%  \\
            & \techWTF{} & 29.67\% & 53.25\% & 13.00\% & 48.03\% & 11.52\% & 24.80\% & 5.45\% & 24.44\% & 16.33\% & 36.63\% \\
            & \techWoA{} & 29.00\% & 52.02\% & 12.33\% & 45.93\% & 10.91\% & 22.15\% & 4.85\% & 23.83\% & 16.00\% & 35.71\% \\
            & \techWoAR{} & 23.33\% & 40.80\% & 10.67\% & 38.65\% & 9.70\% & 18.85\% & 4.24\% & 16.62\% & 14.00\% & 29.27\% \\
            & \textbf{\tech{}} & \textbf{35.33\%} & \textbf{59.62\%} & \textbf{14.33\%} & \textbf{50.99\%} & \textbf{13.33\%} & \textbf{28.67\%} & \textbf{7.88\%} & \textbf{30.16\%} & \textbf{20.67\%} & \textbf{45.58\%}   \\ \midrule
            
            \multirow{5}{*}{Qwen2.5-Coder}  & \techWoPTC{} & 31.33\% & 49.39\% & 13.33\% & 45.30\% & 12.73\% & 25.04\% & 9.09\% & 24.64\% & 16.00\% & 34.01\% \\
            & \techWoT{} & 33.33\% & 55.88\% & 15.00\% & 47.54\% & 13.94\% & 29.73\% & 9.09\% & 26.95\% & 15.67\% & 35.91\%  \\
            & \techWTF{} & 32.67\% & 55.87\% & 14.67\% & 49.33\% & 13.94\% & 30.12\% & 8.48\% & 27.45\% & 15.00\% & 33.77\% \\
            & \techWoA{} & 32.00\% & 55.18\% & 13.67\% & 48.04\% & 13.33\% & 29.85\% & 7.88\% & 26.98\% & 14.33\% & 32.14\% \\
            & \techWoAR{} & 24.00\% & 47.33\% & 11.33\% & 44.13\% & 13.33\% & 29.65\% & 7.27\% & 27.08\% & 13.67\% & 34.06\%  \\
            & \textbf{\tech{}} & \textbf{37.67\%} & \textbf{60.63\%} & \textbf{16.67\%} & \textbf{51.53\%} & \textbf{15.15\%} & \textbf{32.44\%} & \textbf{10.30\%} & \textbf{34.11\%} & \textbf{18.67\%} & \textbf{42.09\%}   \\ \midrule

            \multirow{5}{*}{GPT-4o-mini} & \techWoPTC{} & 51.00\% & 62.31\% & 20.33\% & 56.34\% & 17.58\% & 28.58\% & 13.33\% & 30.02\% & 33.67\% & 46.98\% \\
            & \techWoT{} & 53.00\% & 68.74\% & 20.67\% & 59.95\% & 22.42\% & 39.00\% & 14.55\% & 34.27\% & 34.33\% & 49.37\%  \\
            & \techWTF{} & 52.00\% & 67.76\% & 21.00\% & 60.07\% & 21.82\% & 38.56\% & 15.15\% & 37.07\% & 35.00\% & 49.33\% \\
            & \techWoA{} & 50.33\% & 65.24\% & 20.33\% & 57.46\% & 20.61\% & 37.29\% & 13.33\% & 35.19\% & 34.00\% & 48.07\% \\
            & \techWoAR{} & 42.67\% & 56.53\% & 18.33\% & 52.06\% & 21.82\% & 38.44\% & 12.73\% & 34.28\% & 30.67\% & 46.18\%  \\
            & \textbf{\tech{}} & \textbf{56.33\%} & \textbf{72.98\%} & \textbf{22.67\%} & \textbf{63.55\%} & \textbf{23.64\%} & \textbf{40.65\%} & \textbf{17.58\%} & \textbf{42.10\%} & \textbf{38.67\%} & \textbf{57.34\%}   \\ \midrule

            \multirow{5}{*}{Gemini-1.5-Flash} & \techWoPTC{} & 59.67\% & 70.91\% & 24.67\% & 61.08\% & 41.21\% & 51.51\% & 33.33\% & 45.86\% & 37.00\% & 52.10\% \\
            & \techWoT{} & 61.67\% & 74.14\% & 27.67\% & 63.59\% & 43.03\% & 56.62\% & 34.55\% & 49.60\% & 34.00\% & 52.01\%  \\
            & \techWTF{} & 60.67\% & 71.72\% & 27.33\% & 60.78\% & 41.82\% & 53.93\% & 31.52\% & 43.64\% & 36.33\% & 52.32\% \\
            & \techWoA{} & 59.00\% & 70.22\% & 24.33\% & 59.35\% & 40.61\% & 53.71\% & 30.91\% & 43.08\% & 35.33\% & 51.93\% \\
            & \techWoAR{} & 54.33\% & 66.91\% & 23.00\% & 57.93\% & 38.79\% & 49.37\% & 29.70\% & 42.11\% & 31.00\% & 46.78\%  \\
            & \textbf{\tech{}} & \textbf{65.33\%} & \textbf{78.15\%} & \textbf{29.33\%} & \textbf{65.91\%} & \textbf{44.85\%} & \textbf{56.96\%} & \textbf{36.36\%} & \textbf{54.34\%} & \textbf{38.67\%} & \textbf{58.42\%}   \\ 
            % \midrule
            \bottomrule
        \end{tabular}
    \end{adjustbox}
    \vspace{-4mm}
\end{table*}

\subsection{RQ3: Contribution of Main Components}
\label{subsec:RQ3}

\subsubsection{Variants:}
\tech{} consists of three main components: misaligned specification identification, specification lifting, and specification alignment.
To investigate the contribution of each component, we construct five variants of \tech{} for evaluation.

For the misaligned specification identification component, we utilize both public test cases and LLM-generated test cases to assess the correctness of the initial code. 
To investigate separately the contribution of these two types of test cases, we construct two variants of \tech{}: namely \textbf{\techWoPTC{}} and \textbf{\techWoT{}}, where public test cases and the tester agent are removed, respectively.

For the specification lifting component, existing agent-based code generation techniques typically rely on the coarse-grained execution results of test cases as feedback to repair incorrect code.
In contrast, \tech{} employs specification lifting to obtain finer-grained feedback messages. 
To assess the effectiveness of this component, we replace the specification lifting strategy with a test-execution feedback strategy (introduced in Self-repair~\cite{olausson2024selfrepair}), resulting in a variant of \tech{} called \textbf{\techWTF{}}. 
Specifically, \techWTF{} utilizes test-execution feedback messages as input for the subsequent aligner agent to align the specification.

For the specification alignment component, we evaluate the contributions of both the aligner agent and pre-defined alignment rules by constructing two additional variants of \tech{}: \textbf{\techWoA{}} and \textbf{\techWoAR{}}. 
In \techWoA{}, we remove the aligner agent and randomly select pre-defined alignment rules for specification alignment. 
In \techWoAR{}, we eliminate the pre-defined alignment rules and instead provide a task instruction that guides the aligner agent to directly generate the aligned specification.

\subsubsection{Results:}
Table~\ref{tab:rq3} shows the comparison results between \tech{} and its five variants across all 20 subjects (4 LLMs $\times$ 5 benchmarks) in terms of Pass@1 and AvgPassRatio.
Firstly, \tech{} consistently outperforms both \techWoPTC{} and \techWoT{} in terms of both metrics. 
On average, \tech{} improves 17.59\% and 12.41\% higher Pass@1 than \techWoPTC{} and \techWoT{}, and 25.12\% and 11.84\% higher AvgPassRatio, respectively. 
These results confirm the effectiveness of incorporating both public test cases and LLM-generated test cases in \tech{}. 
Additionally, \techWoT{} demonstrates superior effectiveness over \techWoPTC{}, achieving average improvements of 4.27\% and 10.20\%, respectively. 
This underscores the necessity of our hierarchical criteria (introduced in Section~\ref{subsec:alignment}), which confirms that while public test cases are typically more reliable, LLM-generated test cases may introduce errors. 
Moreover, prior studies~\cite{tian2025fixing,mathews2024tgen,zhang2024paircoder} have suggested that including public test cases within specifications is a common practice. 
However, \techWoPTC{} addresses a more challenging scenario where no public test cases are available. 
The experimental results show that \techWoPTC{} still outperforms all 10 baselines, even those leveraging public test cases, with average improvements of 7.37\%$\sim$48.25\% and 2.17\%$\sim$33.15\% in terms of Pass@1 and AvgPassRatio, respectively. 
This highlights the generalizability of \tech{}, demonstrating its effectiveness even in practical settings where public test cases are unavailable.

Secondly, \tech{} demonstrates superior performance compared to \techWTF{}, achieving average improvements of 14.75\% and 12.95\% in terms of Pass@1 and AvgPassRatio, respectively. 
These results validate the contribution of the specification lifting component and further confirm that it outperforms the test-execution feedback strategy commonly employed by existing agent-based techniques. 
In the future, we aim to further explore the integration of our specification lifting strategy to enhance the effectiveness of existing agent-based code generation techniques.

Thirdly, \tech{} outperforms both \techWoA{} and \techWoAR{} with average improvements of 20.94\% and 34.57\% in terms of Pass@1, and 16.49\% and 28.75\% in terms of AvgPassRatio, respectively. 
These results confirm the necessity of the specification alignment component in \tech{}. 
Moreover, \techWoAR{} demonstrats the weakest performance among all variants, further underscoring the significance of the ten alignment rules derived from software requirements engineering, which contribute the most substantially to the overall effectiveness of \tech{}.

Furthermore, the \textit{Wilcoxon Signed-Rank Test}~\cite{wilcoxon1970critical} at a significance level of 0.05 confirms that all p-values are smaller than \num{1.77e-7}, exhibiting the statistically significant advantage of \tech{} over all variants.
Overall, each of the main components contributes substantially to the overall effectiveness of \tech{}.

% \vspace{-1mm}
\section{Discussion}
\label{sec:discussion}

\subsection{Case Study}
\label{subsec:interpretability}
\begin{figure}[t!]
    \centering
    \includegraphics[width=1.0\linewidth]{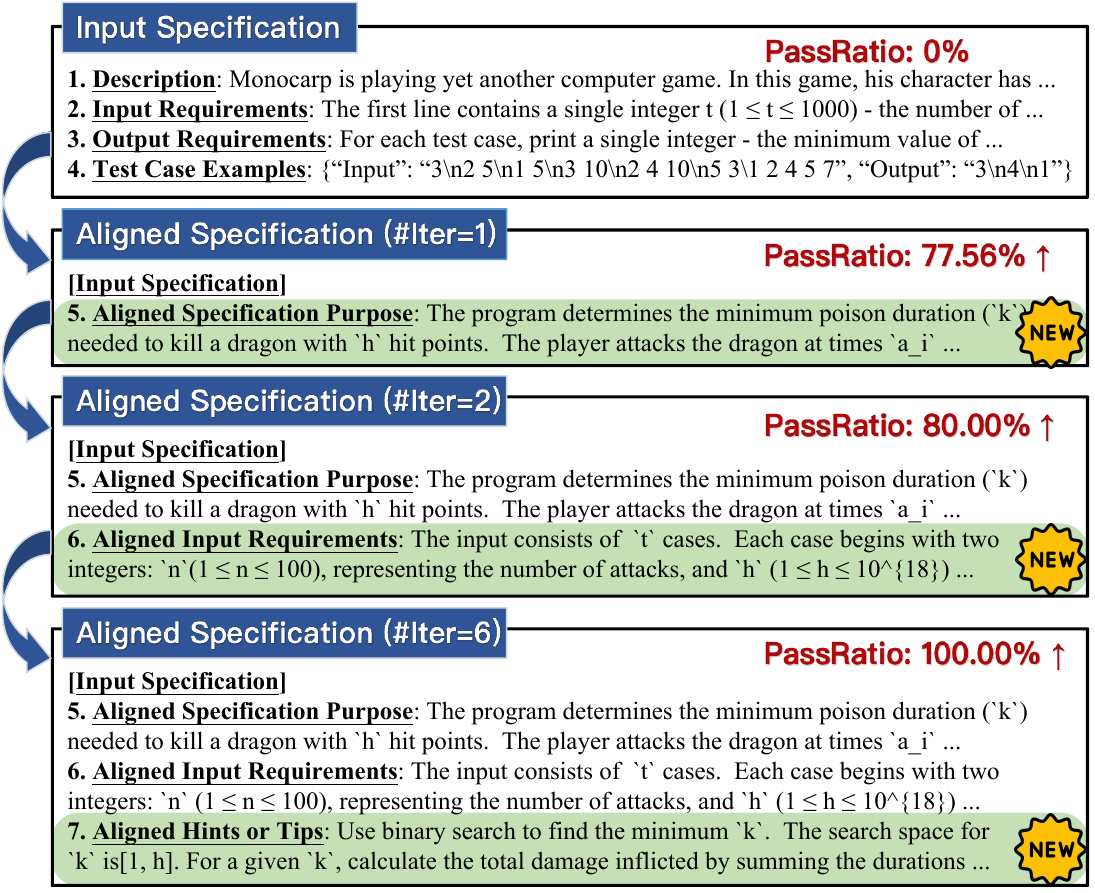}
    \vspace{-4mm}
    \caption{Case study on CodeContests with Gemini-1.5}
    \label{fig:case_study}
    \vspace{-4mm}
\end{figure}

During the alignment process, \tech{} outputs intermediate aligned specifications, enabling systematic analysis of its iterative alignment process. 
We present a case study using CodeContests benchmark and Gemini-1.5 (shown in Figure~\ref{fig:case_study}). 
The quality of the aligned specification is quantitatively evaluated based on the code correctness with private test cases:
(1) Initially, the LLM generates code solely from the input specification, achieving a 0\% pass ratio, indicating that the LLM fails to correctly perceive the specification. 
(2) \tech{} leverages the \textit{specification purpose} rule to generate an aligned specification, improving the pass ratio to 77.56\%, demonstrating that this alignment rule enhances the LLM's perception. Consequently, this alignment rule is retained.
(3) In the next iteration, \tech{} incorporates the \textit{input requirements} rule, increasing the pass ratio to 80.00\%, confirming its contribution.
(4) Subsequent iterations yield no further improvements, leading to the removal of ineffective alignment rules.
(5) By the sixth iteration, \tech{} incorporates the \textit{hints or tips} rule, achieving a 100.00\% pass ratio, indicating that the LLM fully perceives the final aligned specification.
This provides a clear illustration of how \tech{}'s iterative alignment mechanism enhances the LLM perception, ultimately improving code generation performance.

% In-depth analysis of failure cases.
\revision{
To better understand the limitations of \tech{}, we conduct a manual analysis of 60 representative failure cases, sampled across four LLMs and five benchmarks. 
The analysis reveals that most failures are attributed to (1) suboptimal test cases generated by the tester agent (23 cases) and (2) incorrect application of alignment rules by the aligner agent (25 cases). 
A smaller subset (12 cases) involves inaccurate lifted specifications produced by the lifter agent. 
These findings highlight opportunities for further enhancement.
In future work, we plan to enhance the robustness of the tester agent by incorporating specification alignment into the test generation process and introducing test case validation mechanisms. 
For the aligner agent, we intend to develop more effective alignment rules to mitigate specification misalignment.}

\subsection{Influence of Alignment Rules}
\label{subsec:alignment_rules}
\begin{figure}[t!]
    \centering
    \includegraphics[width=1.0\linewidth]{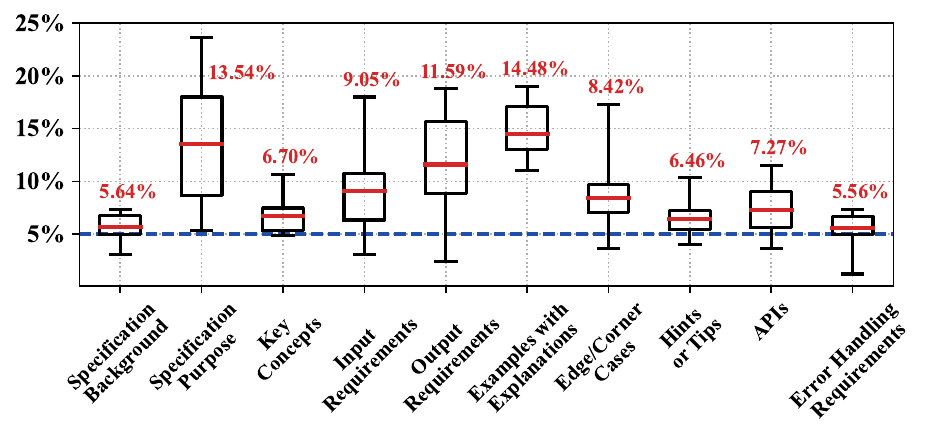}
    \vspace{-4mm}
    \caption{Effectiveness of each alignment rule in \tech{}}
    \label{fig:rule_contribution}
    \vspace{-4mm}
\end{figure}

We further analyze the influence of each alignment rule (as defined in Section~\ref{subsec:alignment}). 
Specifically, for each programming problem, if a specific alignment rule leads to an improvement in the test pass ratio of generated code, it is considered an effective alignment rule for that problem. 
Based on this criterion, we evaluate the contribution of each alignment rule across all programming problems, considering all four LLMs and five benchmarks (as illustrated Figure~\ref{fig:rule_contribution}). 
The results demonstrate that, on average, all alignment rules contribute to aligning more than 5\% of all programming problems, further strengthening their necessity.
Additionally, we observe that the three most effective alignment rules are \textit{Examples with Explanations}, \textit{Specification Purpose}, and \textit{Output Requirements}, which successfully align an average of 14.48\%, 13.54\%, and 11.59\% of all programming problems, respectively. 
This finding suggests that developers should focus on improving these three aspects when writing programming specifications for LLMs in code generation.
In the future, we plan to design additional alignment rules (e.g., performance, security, and deployment requirements) to further enhance \tech{}'s effectiveness in handling more complex problems.

% \vspace{-1mm}
\subsection{Quality of LLM-Generated Test Cases}
\label{subsec:test_quality}
The quality of LLM-generated test cases directly impacts the effectiveness of \tech{}. 
To comprehensively evaluate test case quality, we verify their correctness against the canonical solutions provided in the dataset follow existing research~\cite{huang2023agentcoder}. 
That is, an LLM-generated test case is considered correct if it passes the canonical solution. 
Specifically, the accuracy of LLM-generated test cases is 60.17\%$\sim$72.92\% on average across different LLMs and benchmarks, indicating that a non-negligible proportion of LLM-generated test cases contain errors.
We acknowledge that the presence of incorrectly generated test cases is a general challenge for all LLM-based approaches, primarily due to inherent limitations in LLM performance. 
Moreover, the accuracy of the generated test cases exceeds 50\%, with the proportion of correct test cases surpassing that of incorrect ones. 
Therefore, this facilitates prioritization of correct code, as it exhibits a higher pass ratio on LLM-generated test cases than incorrect code, aligning with the voting principle. 
Our ablation study on \techWoPTC{} further confirms that LLM-generated test cases significantly enhance the overall effectiveness of \tech{}, providing empirical evidence of their contribution.
Similar findings have been reported in CodeT~\cite{chen2022codet} and FlowGen~\cite{lin2025flowgen}.
To further relieve this challenge, we introduce a hierarchical pass ratio criterion (in Section~\ref{subsec:alignment}) that prioritizes the results of reliable public test cases. 
This helps mitigate the negative impact of erroneous LLM-generated test cases to some extent.
In future work, we aim to develop more advanced test case generation techniques, such as incorporating type checking and test coverage to guide LLM-based test generation, further improving the effectiveness of \tech{}.

\subsection{Scalability of DSL}
\label{subsec:dsl_scalability}
\revision{
While our current evaluation focuses on structured programming tasks, the DSL design in our \tech{} is designed to be modular and extensible. 
The core principle of \tech{}, which involves abstracting LLM-generated code into a structured and semantically rich representation to improve alignment with user intent, is broadly applicable beyond competitive programming.
The current schema builds on established foundations in requirements engineering~\cite{greenspan1994formal,glinz2000problems} and aligns with \textit{IEEE specification standard}~\cite{doe2011recommended}, ensuring wide applicability across diverse code generation scenarios. 
It can be extended to incorporate domain-specific ingredients (e.g., UI layout constraints in front-end development, data schemas in analytics) without modifying the overall alignment pipeline. 
In future work, we plan to investigate the use of advanced LLMs to generate DSL schemas tailored to new domains, with refinement via expert knowledge.}
% \vspace{-1mm}
\section{Threats and Validity}
\label{sec:threats}
The threat to \textit{construct} validity mainly lies in the inherent randomness involved in LLMs.
On one hand, we conduct a large-scale study, and the consistency of our findings across all subjects helps mitigate this threat.
On the other hand, we repeat \tech{}'s experiments three times across all 20 subjects (4 LLMs $\times$ 5 benchmarks).
Notably, the standard derivations are only 0.008 and 0.022 in terms of Pass@1 and AvgPassRatio, respectively, demonstrating the robustness of our conclusions to a large extent.
Furthermore, a \textit{Wilcoxon Signed-Rank Test}~\cite{wilcoxon1970critical} at the significance level of 0.05 confirms that all p-values exceed 0.55, indicating no statistically significant differences across the three experimental results.
This further strengthens the reliability of our results.

The threat to \textit{external} validity mainly lies in the used subjects. 
To mitigate this, we carefully select a diverse set of benchmarks, metrics, baselines, and LLMs. 
Following prior studies~\cite{islam2024mapcoder,jiang2024self,tian2025fixing}, we utilize five widely-used benchmarks in code generation and employ two metrics to assess code correctness.
Additionally, we compare \tech{} with ten popular or state-of-the-art baselines on four advanced LLMs, ensuring a comprehensive evaluation.
In the future, we will extend our evaluation to a broader set of benchmarks and LLMs to further assess \tech{}'s generalizability.
% \vspace{-1mm}
\section{Related Work}
\label{sec:related}
Code generation is a critical task in software engineering and has garnered significant attention in recent years~\cite{zhang2024paircoder,li2025structured,codeforces2023,tianleam++}. 
Among the various approaches, prompt-based and agent-based techniques are the two most widely-adopted paradigms.

\textbf{Prompt-based techniques} enhance code generation performance of LLMs by designing carefully-crafted prompts that can be applied in a plug-and-play manner. 
Based on their design principles and application stages, these techniques can be categorized into three types:
(1) Pre-generation techniques (e.g., Self-planning~\cite{jiang2024self} and SCoT~\cite{li2025structured}) aim to guide LLMs in generating intermediate reasoning steps, thereby improving the effectiveness of code generation.
(2) Post-generation techniques (e.g., Self-debugging~\cite{chen2024teaching}, Self-edit~\cite{zhang2023self}, and Self-repair~\cite{olausson2024selfrepair}) leverage error messages from test execution to enable LLMs to refine and correct erroneous code.
(3) Hybrid techniques (e.g., $\mu$Fix~\cite{tian2025fixing}) integrate the advantages of both pre-generation and post-generation approaches, leveraging their synergy to further enhance code generation performance of LLMs.

\textbf{Agent-based techniques} improve the code generation performance of LLMs by simulating human collaborative software development processes, where multiple agents interact to accomplish tasks such as task planning, code generation, and testing.
Each agent is assigned a specialized role, and their collaborative interactions optimize different stages of the code generation process.
Representative agent-based techniques, including Self-collaboration~\cite{dong2024selfcollaboration}, AgentCoder~\cite{huang2023agentcoder}, and PairCoder~\cite{zhang2024paircoder} (introduced in Section~\ref{sec:introduction} and ~\ref{subsec:baselines}), highlight the importance of role definition and multi-agent coordination in improving code generation performance of LLMs.

Unlike existing techniques, \tech{} introduces a novel perspective to enhance the code generation performance of LLMs through specification alignment. 
Our experimental results demonstrate that \tech{} significantly outperforms state-of-the-art prompt-based and agent-based code generation techniques in terms of effectiveness. 
Notably, \tech{} is orthogonal to existing prompt-based and agent-based techniques to some extent. 
Specifically, \tech{} can serve as a pre-processing module, aligning requirement specifications before they are provided to other techniques. 
In future work, we plan to explore the integration of \tech{} with existing techniques to further enhance their performance.

\section{Conclusion}
\label{sec:conclusion}
In this work, we draw inspiration from software requirements engineering practices and propose a novel perspective for enhancing the code generation performance of LLMs through specification alignment.
Based on this perspective, we design \tech{}, which consists of identifying misaligned input specifications, lifting LLM-perceived specifications, and aligning them to further generate the correct code solution.
To evaluate the effectiveness of \tech{}, we conduct comprehensive experiments on four advanced LLMs and five challenging benchmarks. 
Experimental results demonstrate that \tech{} consistently outperforms state-of-the-art prompt-based and agent-based code generation techniques across multiple evaluation metrics, highlighting its superiority in enhancing the code generation performance of LLMs.

\begin{acks}
\revision{
This work is supported by the National Natural Science Foundation of China (Grant No. 62322208).}
\end{acks}

\balance
\bibliographystyle{ACM-Reference-Format}
\bibliography{reference}

\end{document}